\documentstyle[bo99,epsfig]{article}

\newcommand{\ros}{{\sl ROSAT }}
\newcommand{\G}{$\Gamma_{\rm x}$ }

\title{ An optical and X-ray study of the pecu-\\[3mm]
        \mbox{   }~~~liar narrow-line quasar QSO\,0117$-$2837  }

\author{ Stefanie Komossa$^{1}$, D. Grupe$^{1}$, V. Burwitz$^{1}$, M. Janek$^{2}$ }
\affil{$^{1}$\,Max-Planck-Institut f\"ur extraterrestrische Physik,
   Giessenbachstr., 85748 Garching, Germany; skomossa@xray.mpe.mpg.de,  
$^{2}$\,3063 Obernkirchen, Germany}

\begin{document}

\maketitle

\vspace*{-8.3cm}
\begin{verbatim}
Contribution to the proceedings of the Bologna workshop on
`X-ray Astronomy 1999' (Sept 6-10, 1999); 
to appear in Astrophys. Letters and Communications
Preprint available at http://www.xray.mpe.mpg.de/~skomossa
\end{verbatim}
\vspace*{6.4cm}

\begin{abstract}
We present an optical and X-ray study of the  
quasar QSO\,0117-2837.
It exhibits an interesting combination
of optical and X-ray properties: 
Despite its extremely steep observed X-ray spectrum 
(\G $\simeq -4$ when fit by a simple powerlaw),
its Balmer lines are fairly broad.  
A two-component Gaussian fit to H$\beta$ yields
FWHM$_{\rm H\beta, broad} \simeq 4000$ km/s,
and places QSO\,0117-2837 in the `zone of avoidance'
in the FWHM$_{\rm H\beta}$--\G diagram.   
A time variability analysis shows that 
QSO\,0117-2837 is another case of Narrow-line Sy\,1 galaxy 
(NLSy1 hereafter) which
does {\em not} show any X-ray variability during the observation.
The results are discussed in view of the 
NLSy1 character of QSO\,0117-2837.

\end{abstract}

\section{ Introduction}

QSO\,0117-2837 (1E 0117.2-2837) was discovered as an X-ray source by
{\sl Einstein} and is
at a redshift of $z$=0.347 (Stocke et al. 1991).
It is serendipituously located in a \ros PSPC pointing.
Its X-ray spectrum is extremely steep, as was 
briefly noted by Schwartz et al. (1993).
We present here the first detailed analysis of the \ros~observations
of this AGN (see Komossa \& Fink 1997b for first results, and
Komossa \& Meerschweinchen 2000 for the complete ones), 
and a discussion of a high-quality optical spectrum.

\section{ Optical properties} 

We have obtained an optical spectrum of QSO\,0117-2837 with 
the ESO\,1.52\,m telescope at LaSilla in September 1995. 
The spectrum covers the wavelength range 3900-7800\AA~with a 
resolution of 6\AA.   
The optical spectrum reveals several signs of a 
NLSy1 galaxy (we do not distinguish between NL Seyferts 
and NL quasars, here): 
weak [OIII]$\lambda$5007 emission and strong
FeII complexes (Fig. 1a).
After subtraction of the FeII spectrum (see Grupe et al. 1999 for details)
we derive FWHM${_{\rm H\beta}}$=2100$\pm{100}$ km/s,
FWHM${_{\rm [OIII]}}$=820$\pm{150}$ km/s
(based upon single-component Gaussian fits to the emission
lines),
and [OIII]/H$\beta$=0.056.        
Alternatively, H$\beta$ was fit with a two-component Gaussian
in which case we obtain FWHM${_{\rm H\beta,broad}} \simeq$  4000 km/s 
and FWHM${_{\rm H\beta,narrow}}$ $\simeq$ 1100 km/s. 
Our spectrum also covers the important region around
[OII]$\lambda$3727, 
which provides an important
discriminant between different models to account for
the excitation of the NLR in NLSy1s (Komossa \& Janek 1999). 
 [OII] is not detected,
implying [OII] to be much weaker than [OIII].    

\section {X-ray properties }

The {\sl ROSAT} PSPC observation of QSO\,0117-2837 was performed 
in Dec. 1991 with an exposure time of 4.5 ksec. 
The source is detected with a countrate of 0.44 cts/s
(see Komossa \& Meerschweinchen 2000 for further details
on X-ray data reductions carried out).  
\vskip0.2cm

\noindent {\it{Spectral analysis.}}
The \ros X-ray spectrum of QSO\,0117-2837 is extremely steep.
Three models provide a successful spectral fit; we discuss each in turn:

(i) When the X-ray spectrum is fit by a single powerlaw
with Galactic cold absorption
of $N_{\rm Gal} = 1.65\,10^{20}$ cm$^{-2}$,
we derive a photon index \G $\simeq -3.6\pm{0.1}$
($-4.3\pm{0.4}$, if $N_{\rm H}$ is treated as free parameter).
The overall quality of the fit is good ($\chi{^{2}}_{\rm red} = 0.8$), but 
slight systematic
residuals remain (Fig. 1b).

(ii) A successful alternative description is a {\em warm-absorbed} flat powerlaw
of fixed canonical index \G = --1.9. 
The warm absorber models are based on Ferland's (1993) code {\em Cloudy},
and the model assumptions and calculations are described in more detail
in Komossa \& Fink (e.g., 1997a). 
We find a very large column density
of the warm absorber, and the contribution
of emission and reflection is no longer negligible; there is also some
contribution to Fe K$\alpha$.
For the pure absorption model, the best-fit values
for ionization parameter and warm column density are $\log U \simeq 0.8$,
and $\log N_{\rm w} \simeq 23.6$
($N_{\rm H}$ is now consistent with the  Galactic value when treated as free parameter),
with $\chi{^{2}}_{\rm red}$=0.7.
Including the contribution of emission
and reflection for 50\% covering of the warm material as calculated with {\em Cloudy} gives
$\log N_{\rm w} \simeq 23.8$ ($\chi{^{2}}_{\rm red}$ = 0.65).
Several strong EUV emission lines are predicted to arise from the warm material
(e.g., FeXXI$\lambda$2304/H$\beta_{\rm wa}$ = 10,
NeVIII$\lambda$774/H$\beta_{\rm wa}$ = 9, and
FeXXII$\lambda$846/H$\beta_{\rm wa}$ = 113).
No absorption from CIV and NV is expected to show up in the UV. 
Both elements are more highly
ionized.

(iii) Thirdly, the spectrum can be fit with a flat powerlaw
(\G fixed to --1.9) plus soft excess which was 
parameterized as black body, or the standard accretion-disk emission
model available in EXSAS.
We find $kT_{\rm bb} \simeq 0.10$ keV for the black body description
($\chi{^{2}}_{\rm red}$=0.7). 
Using the accretion disk description, and fixing the black hole mass
to $M_{\rm BH}=0.6\,10^4 M_{\odot}$ yields ${\dot M}\over{{\dot M_{\rm edd}}}$ = 0.6.
\vskip0.2cm

\noindent {\em{Temporal analysis.}} 
An analysis of the temporal variability reveals {\em constant} source
flux within the
1$\sigma$ error during the observation.
This makes QSO\,0117-2837 another example of a NLSy1 with
constant X-ray flux during the time-interval of observation,
showing that {\em not all} NLSy1s are characterized by permanent 
rapid variability (see also the discussion and examples 
in Wisotzki \& Bade 1997).

\begin{figure}
{\psfig{file=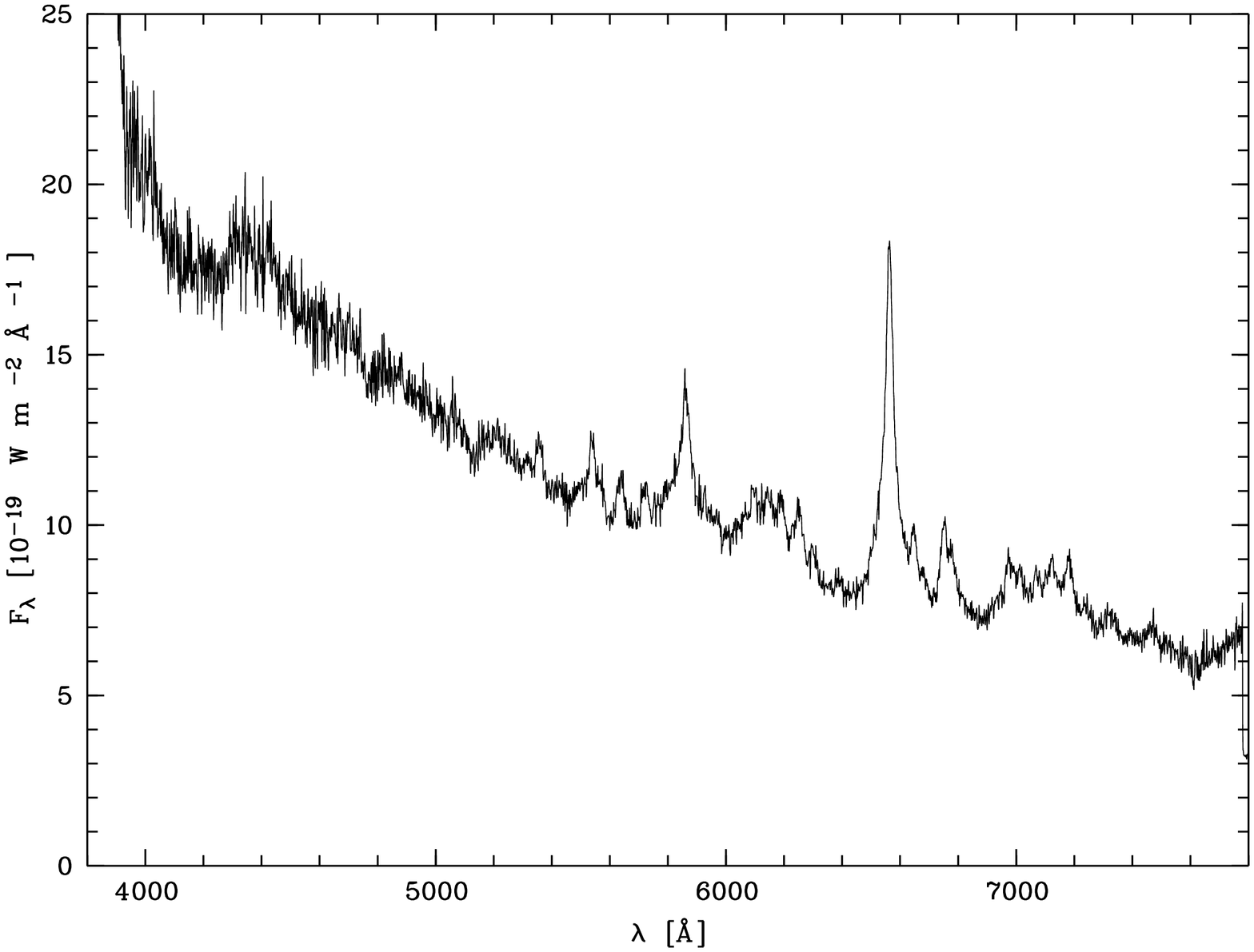,width=5.4cm,angle=-90,clip=}}
\vskip-6.3cm
\hspace*{7.9cm} 
\vbox{\psfig{figure=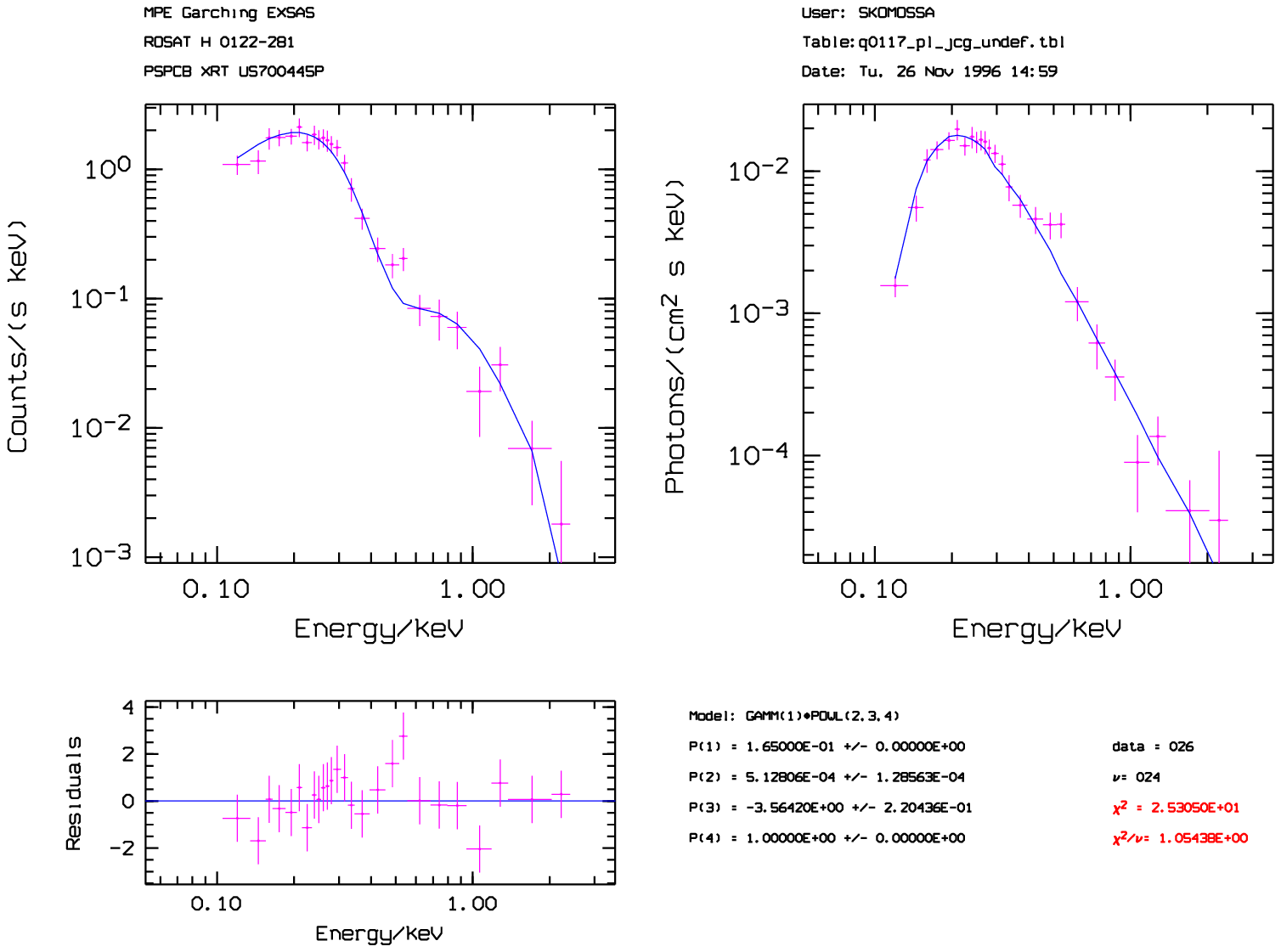,width=4.4cm,height=5.5cm,%
      bbllx=2.1cm,bblly=1.1cm,bburx=10.1cm,bbury=11.7cm,clip=}}\par
\vskip-0.5cm
\hspace*{7.9cm}
\vbox{\psfig{figure=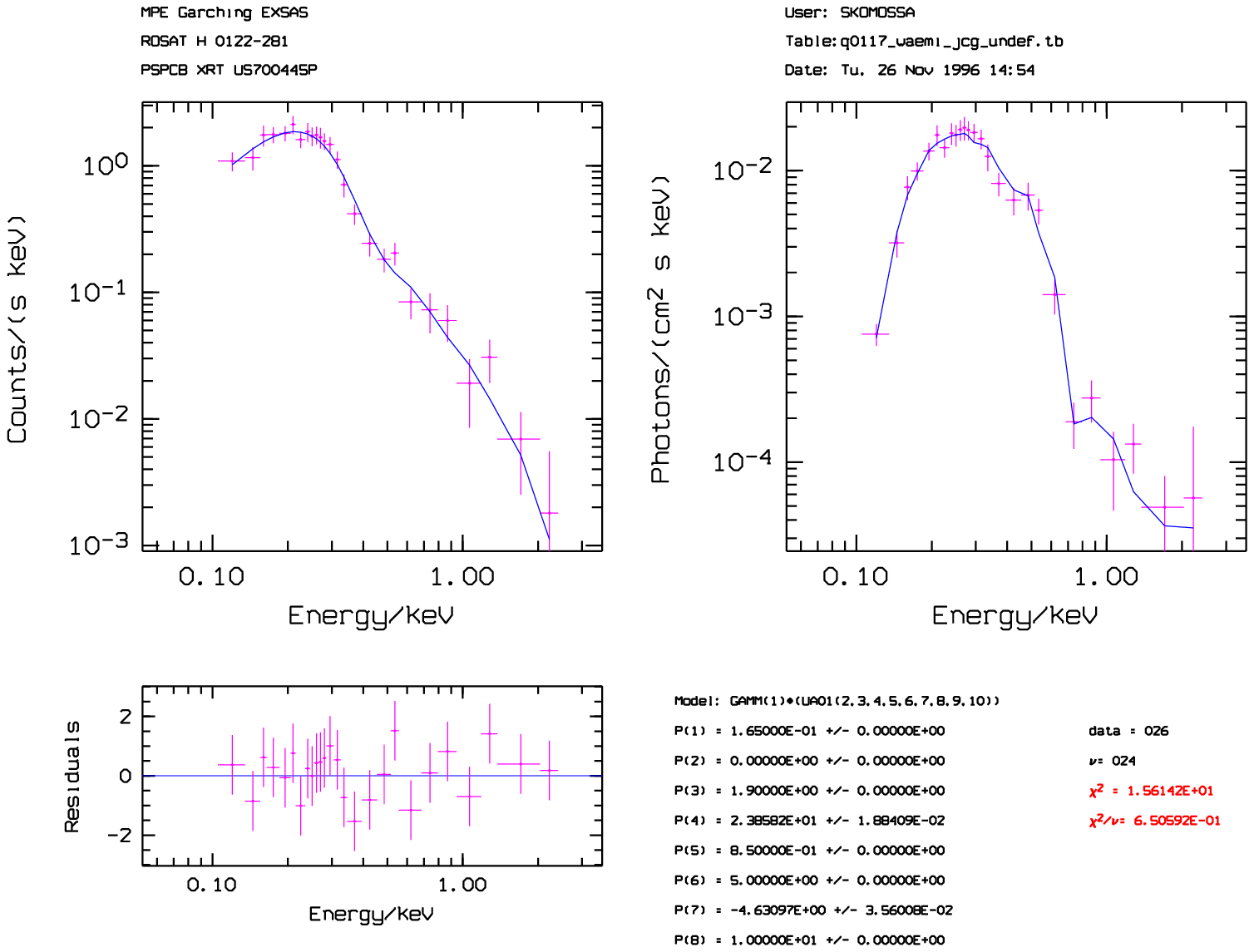,width=4.4cm,height=1.45cm,%
          bbllx=2.1cm,bblly=1.98cm,bburx=10.1cm,bbury=4.5cm,clip=}}\par
\vspace*{-0.2cm} 
\caption[]{{{\bf {(a}}, left{\bf{)}} Optical spectrum of QSO\,0117-2837.
\bf {(b)}} \ros X-ray spectrum of QSO\,0117-2837.
The first panel shows the observed X-ray spectrum (crosses) and best-fit
powerlaw (solid line). The second panels shows the residuals
for this model description whereas the third panel displays the
residuals after fitting a warm absorber model.  }
\end{figure}

\section{Discussion }

Given the rather large width of H$\beta$, 
the X-ray spectrum of QSO\,0117-2837 is exceptionally steep;
among the steepest observed in NLSy1s. 
In fact, whereas there is a very large scatter in the X-ray
spectral steepness of NLSy1s, 
with several as flat as 'normal'
Seyferts (e.g., Xu et al. 1999), broad line objects always
tend to show flat X-ray spectra (e.g., Boller et al. 1996, Grupe et al. 1999);
the corresponding region in the FWHM$_{\rm H\beta}$--\G diagram
is occasionally referred to as `zone of avoidance'. 
QSO\,0117-2837 appears to be an important transition object:  
Depending on which representation of the line profile 
is chosen -- a one-component Gaussian with FWHM$_{\rm H\beta}$=2100 km/s,
or a two component Gaussian which gives a better fit and  
FWHM$_{\rm H\beta,broad}$=4000 km/s -- QSO\,0117-2837 is placed 
at the border of the `zone of avoidance', or inside it, respectively.   

The origin of the steep observed X-ray spectra of NLSy1 galaxies
and the relation to their optical properties 
is still not well understood. 
In particular, it is interesting to point out the following: 
Whereas among the early suggestions to explain the X-ray spectral
steepness of NLSy1s was the presence of a strong soft excess, in
analogy to `normal' Seyferts and quasars that were believed to show
a soft excess, 
there is now growing evidence that many soft excesses in Seyferts were in fact
mimicked by the presence of warm absorbers.
On the other hand, the soft excesses in NLSy1s, although originally
inferred in analogy to Seyferts, turn out to be real, as judged
by recent {\sl ASCA} and {\sl SAX} observations (e.g., Vaughan et al. 1999).
Often, it turns out, several components are simultaneously
present that contribute to the spectral steepness in
NLSy1s: A steeper-than-usual powerlaw, a soft excess {\em and}
a warm absorber 
pointing to the spectral complexity in these objects
(see, e.g., the discussion in Komossa \& Fink 1997a,b).

In the case of QSO\,0117-2837 the limited \ros spectral resolution
does not allow us to distinguish between models (i)-(iii) of Sect. 3.
It still allows to determine the maximal possible contribution
of each component (i)-(iii).   
In fact, the rather large inferred column density $N_{\rm w}$
of the warm absorber (model ii) suggests that further mechanisms contribute to,
or dominate, the
X-ray steepness.
Given the very steep rise towards the blue of QSO\,0117-2837's optical
spectrum (with $\alpha_{\rm opt,x}$=0.0, Grupe et al. 1998), 
it is tempting to speculate that a giant soft-excess
dominates the optical-to-X-ray spectrum. We strongly caution,
though, that simultaneous optical-X-ray variability studies
in other Seyferts and NLSy1s (e.g., Done et al. 1995)
do {\em not} favor a common origin of X-ray and
optical components. Secondly, one giant optical-to-X-ray
bump seems to be inconsistent with the finding of 
Rodriguez-Pascual et al. (1997) that NLSy1s tend to be underluminous
in the UV. The possibility of an {\em indirect} relation between
optical and X-ray component remains.     

Given the abundant presence of warm absorbers in `normal Seyferts', and the
additional trend 
that warm absorbers are more abundant in
FeII-strong objects (Wang et al. 1996), combined with 
QSO\,0117-2837's large FWHM$_{\rm H\beta}$ and steep X-ray spectrum,
it is
likely that both, a soft excess and a warm absorber contribute to 
its spectral steepness in the \ros band. 
Its peculiar optical--X-ray properties make the quasar 
 a good target for (a) follow-up 
X-ray spectral observations with, e.g., {\sl XMM},
and (b) high-resolution optical studies of the H$\beta$ profile.

\vskip0.15cm
{\small \noindent This and related papers can be retrieved from our webpage at \\ 
http://www.xray.mpe.mpg.de/$\sim$skomossa/ }

\end{document}